\documentclass[conference]{IEEEtran}
\IEEEoverridecommandlockouts

\usepackage[hyphens]{url}
\usepackage[hidelinks]{hyperref}
\usepackage{tabularray}
\usepackage{xcolor}
\usepackage{orcidlink}

\usepackage{xspace}

\newcommand{\boldit}[1]{\textbf{\emph{#1}}\xspace}

\newcommand{\urlfootnote}[1]{\footnote{\urltt{#1}}\xspace}
\newcommand{\urlttlight}[1]{\url{#1}}
\newcommand{\urltt}[1]{\textbf{\url{#1}}}

\newcommand{\ie}{\emph{i.e.,}\xspace}
\newcommand{\eg}{\emph{e.g.,}\xspace}

\newcommand{\etal}{\emph{et~al.}\xspace}

\newcommand{\ritgard}{{\sc Ritgard}\xspace}
\newcommand{\replipackage}{\urltt{https://doi.org/10.6084/m9.figshare.30444704}}

\definecolor{tablelightgray}{gray}{0.87}
\definecolor{tabledarkgray}{gray}{0.2}

\usepackage{tikz}

\newcommand\encircle[1]{%
  \tikz[baseline=(X.base)] 
    \node (X) [draw, shape=circle, inner sep=-1pt] {\strut #1};}

\newcommand{\topic}[1]{``\emph{#1}''\xspace}

\newcommand{\insightBox}[1]{\vspace{.5em}\noindent\framebox[\columnwidth][c]{\parbox[b]{0.95\columnwidth}{ #1 }}\vspace{.5em}}

\usepackage{etoolbox}
\AtBeginEnvironment{thebibliography}{%
   \interlinepenalty10000%
}

\newcommand{\Description}[1]{}

\usepackage{doi}

\usepackage{csquotes}

\renewcommand{\mkbegdispquote}[2]{\itshape\openautoquote}

\usepackage{enumitem}

\begin{document}
\title{\huge T(r)opical Islands:\\ Visualizing \& Understanding Socio-Technical Artifacts}


\author{%
    \IEEEauthorblockN{%
        Adam Štěpánek\IEEEauthorrefmark{1}\orcidlink{0009-0008-9388-2546},\,%
        Marco Raglianti\IEEEauthorrefmark{2}\orcidlink{0000-0002-6878-5604},\,%
        Vít Rusňák\IEEEauthorrefmark{1}\orcidlink{0000-0003-1493-2194},\,%
        Jan Byška\IEEEauthorrefmark{1}\IEEEauthorrefmark{3}\orcidlink{0000-0001-9483-7562},\,%
        Barbora Kozlíková\IEEEauthorrefmark{1}\orcidlink{0000-0003-0045-0872},\,%
        Michele Lanza\IEEEauthorrefmark{2}\orcidlink{0000-0003-4391-0197}}
    \IEEEauthorblockA{%
        \IEEEauthorrefmark{1}Visitlab, Masaryk University, Brno, Czech Republic}
    \IEEEauthorblockA{%
        \IEEEauthorrefmark{2}REVEAL @ Software Institute, USI, Lugano, Switzerland}
    \IEEEauthorblockA{%
        \IEEEauthorrefmark{3}VisGroup, University of Bergen, Bergen, Norway}
}

\maketitle

\begin{abstract}

Projects hosted on collaborative software development platforms, such as GitHub, include many non-code artifacts documenting the project's lifecycle, with its challenges, plans, design, and even community. These socio-technical artifacts include, for example, bug reports, feature requests, and forum posts, offering a useful prospect on the project's evolution. However, these artifacts are dispersed over multiple communication channels and written in natural language, making their analysis difficult, as they are fragmented and with considerable noise. 

We present a 3D visualization approach mapping topics found across a project's socio-technical artifacts onto vegetation-covered islands, where the individual artifacts are depicted as trees of various types. The topic islands rise out of the ocean as they become discussed, to sink again when they are no longer so. We built a prototype implementing the entire visualization pipeline, from data mining to interactive rendering, leveraging machine learning techniques to cluster the artifacts and extract their topics. We present, through several case studies, the insights that our approach elicits about discussions of development topics throughout a project's history. The user study we conducted (N~=~34) further strengthens our conclusions about its suitability for understanding socio-technical artifacts and their evolution.

\end{abstract}

\begin{IEEEkeywords}
software visualization, topic modeling, developer conversations, socio-technical artifacts, GitHub
\end{IEEEkeywords}



 
\section{Introduction} \label{sec:introduction}

GitHub is the dominant platform for collaborative open-source software development~\cite{octoverse_2024}. While the platform's main purpose is to host Git repositories, it offers a plethora of additional features for managing software projects.

\textit{Issues}, \textit{Pull Requests} (PRs), and \textit{Discussions} are commonly used non-code artifacts documenting the project's evolution~\cite{tantisuwankul_2019,hata_2021}. Issues act as both bug reports and feature requests. PRs contain code changes and their description, waiting to be reviewed, improved, and possibly merged into the project. Finally, Discussions are free-form conversations, brainstormings, or Q\&As, similar to those found on internet forums. All of these artifacts contribute to the project's documentation~\cite{raglianti_2023,venigalla_2024} and chronicle the development of its features and community participation. As the nature of these artifacts encompasses both social and technical aspects, they are termed as \textit{Socio-Technical Artifacts}~\cite{gregor_2013} (STAs) in this paper.

STAs are a key instrument in understanding software evolution, since they describe desiderata, major problems, rationale of changes, and plans for the future of the project~\cite{alkadhi_2017}. However, these artifacts are by their nature unstructured, written in a natural language (therefore with inherent ambiguities and noise), and dispersed over multiple communication channels.

To further complicate matters, these communication channels are intended for different stakeholders. Typically, Discussions are intended for the whole community, PRs mainly for developers, and Issues have different uses in different projects, with a mix of developers and tech-oriented users alike~\cite{github_2026a,github_2026b}.

The topics manifesting in each channel also change over time. For example, what has been the focus of discourse among core developers at the beginning of the project might no longer be relevant. The challenge is in extracting useful insights from these information and documentation sources.

\begin{figure*}[ht]
    \includegraphics[width=\textwidth]{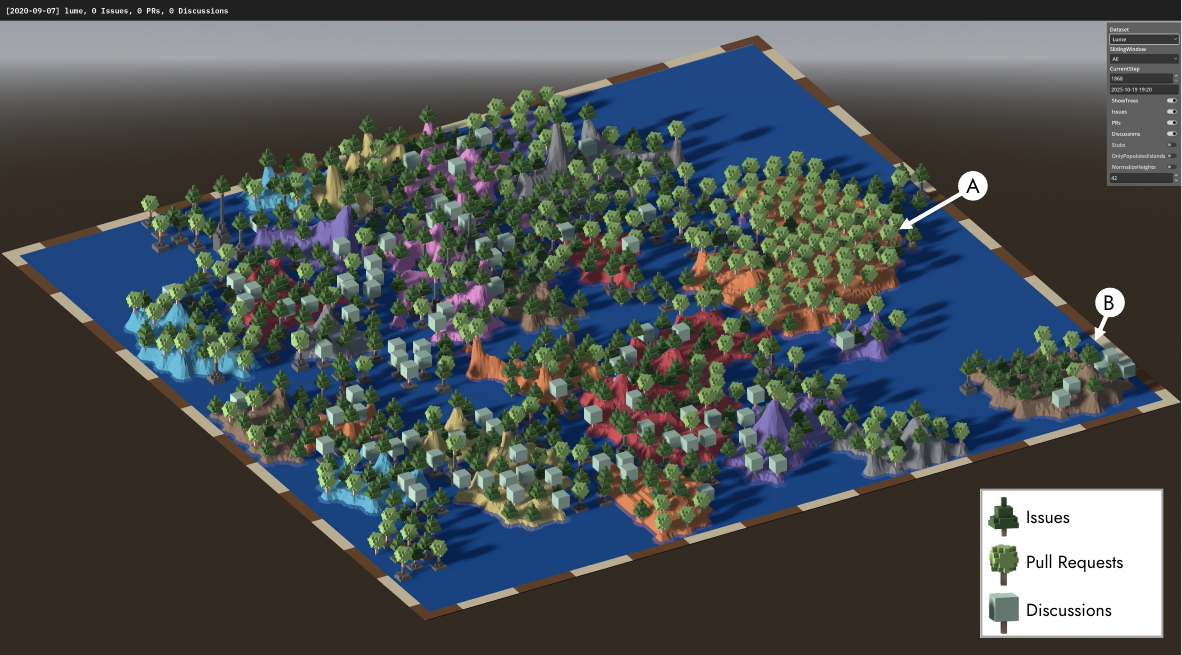}
    \caption{\textit{Lume} in \ritgard. Trees represent Issues, PRs, and Discussions. Islands, such as (A, B), group them by topic.}
    \label{fig:teaser}
    \Description{An archipelago of colorful low-poly islands representing topics found among the socio-technical artifacts of the \texttt{lumeland/lume} GitHub repository. Each island is covered with trees depicting Issues, PRs, or Discussions.}
\end{figure*}

We present an approach to visualize GitHub STAs. We analyze their evolution over time at different granularities. Using machine learning (ML) techniques, including a large language model (LLM), we cluster STAs and identify conversation topics. Each topic is visualized as an island, covered with trees representing STAs revolving around the topic. The liveliness of each topic is depicted by the terrain (\ie landmass) of the island. The more active a topic, the higher the island's terrain. For example, \autoref{fig:teaser} shows the topics found in the repository of \textit{Lume}, a static website generator. Islands rise out of the ocean if their corresponding topic is active, or sink back when the topic is not discussed anymore.

Our approach aids in understanding the project's development across multiple areas of interest, revealing problematic subjects and the conversation hotspots of its community. We developed \ritgard, a tool for the data mining, processing, and visualization of STAs. In three case studies and a user study with 34 participants, we illustrate the advantages of this approach. We show how the visualization reflects the projects' characteristics, through the shape of the landscape and its vegetation, in an intuitive and engaging way.

 \vspace{-1mm}
\section{Mining and Visualizing GitHub STAs} \label{sec:github_stas}

Socio-Technical Artifact is a term used in design science, connected to information systems and software engineering~\cite{gregor_2013,drechsler_2015,weigand_2021,runeson_2020}. STAs are objects with a strong interplay of social and technical aspects in their usage and design~\cite{bijker_1987}. For example, Storey \etal developed a ``Who-What-How''~\cite{storey_2020} framework to assess how software engineering research encompasses human and social aspects. This research field naturally includes collaborative development platforms as generators and repositories of STAs. For instance, Bouraffa \etal studied the social effects of code suggestions in PRs and their impact on the non-code, social-oriented artifacts~\cite{bouraffa_2025}.

Although GitHub STAs are studied and recognized as documentation sources~\cite{raglianti_2023}, to the best of our knowledge, a holistic visual representation of these artifacts and the emerging landscape they generate is still missing. In particular, there is no existing representation effectively capturing the time dimension, key in evolutionary analysis.

\begin{figure*}[ht]
    \centering
    \includegraphics[width=0.99\textwidth]{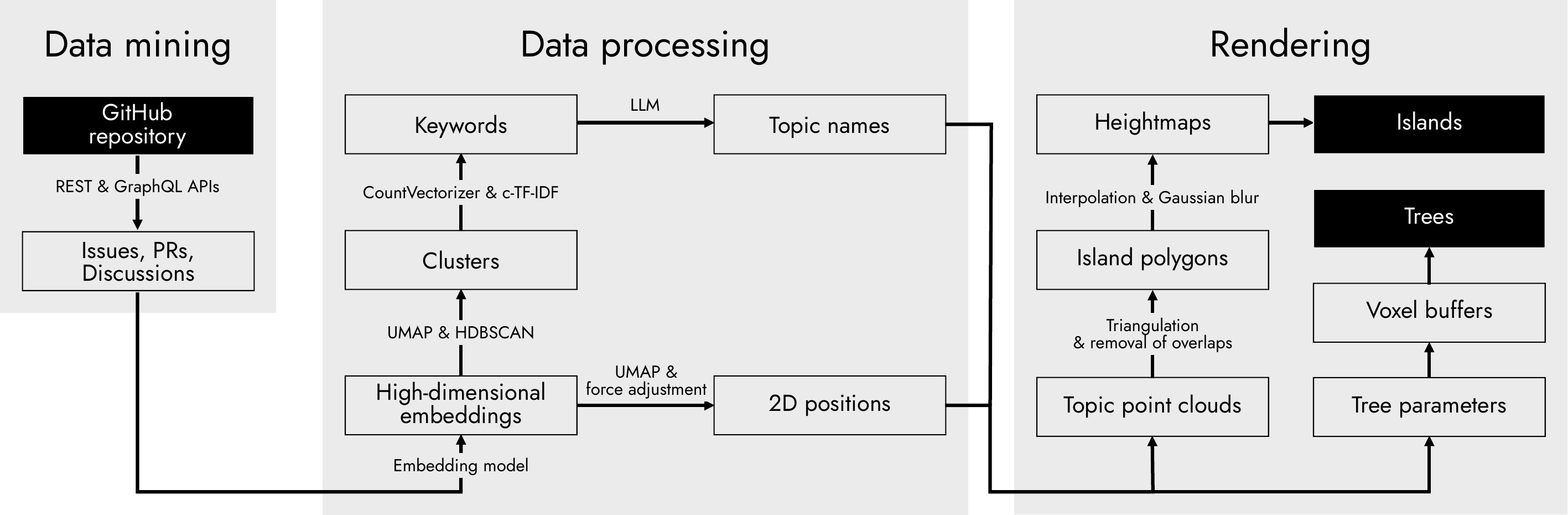}
    \caption{Overview of the visualization approach, as implemented in \ritgard.}
    \label{fig:pipeline}
\end{figure*}

In \autoref{fig:pipeline}, we show an overview of our approach. STAs are mined from GitHub through its Representational State Transfer (REST)\urlfootnote{https://docs.github.com/en/rest?apiVersion=2022-11-28} and Graph Query Language (GraphQL)\urlfootnote{https://docs.github.com/en/graphql} Application Programming Interfaces (APIs). Then, their textual contents are pre-processed, run through an embedding model to generate high-dimensional embeddings, and clustered by semantic similarity. Afterward, we use an LLM to generate a short descriptive label for each cluster---the topic island's name. Finally, we use the Godot\urlfootnote{https://godotengine.org} game engine to render the topics and STAs as islands and trees, respectively.


\subsection{Pre-Processing and Topic Modeling} \label{sec:preprocessing}

In the data pre-processing steps, we convert the mined STAs into plain text. First, the  title, body, and comments of each STA are concatenated. Then, we remove links, figures, tables, code snippets, and Markdown syntax. Finally, the text is prepended with the STA's labels in square brackets. Optionally, some parts (\eg comments) may be omitted so that the final textual representation fits the size of the embedding model's context.

In the data processing stage (\autoref{fig:pipeline}, Data processing), we model the topics using the BERTopic library by Grootendorst~\cite{grootendorst_2022}. For each pre-processed STA, a \textit{text embedding} is computed using the \textit{Qwen3-Embedding-8B} model~\cite{zhang_2025}. Since we need to identify STAs concerning the same topic, we choose the best-performing model from the multilingual Massive Text Embedding Benchmark (MTEB) leaderboard~\cite{mteb} in the semantic textual similarity task as of October 2025.

We reduce the dimensionality of the high-dimensional embeddings using the Uniform Manifold Approximation and Projection for Dimension Reduction (UMAP) algorithm~\cite{mcinnes_2018}. The reduced embeddings are then clustered using HDBSCAN*~\cite{mcinnes_2017}. For each cluster, we compute a bag-of-words representation of word frequency within the included STAs. Then, we remove stop words and pick the most representative keywords with a class-based TF-IDF algorithm~\cite{grootendorst_2022}.

As a final processing step, we compose a prompt for each cluster asking for a topic name. The prompt follows a pre-defined template (included in the \hyperref[replipackage]{replication package}) and contains the titles of the most representative documents (at least four of them), keywords of the topic, keywords of the GitHub repository set by its maintainers, and two examples of expected outputs (\ie for few-shot learning). This prompt is then sent to an LLM, and the resulting topic name is associated with the cluster to be shown in the visualization. We selected OpenAI's \textit{GPT-OSS-B120}~\cite{openai_2025} model for the task, since it was the latest and largest general-purpose LLM available on our research infrastructure.

\subsection{Turning GitHub STAs into Verdant Islands} \label{sec:islands}

To visualize the STAs and their topics, we build upon the \textit{island metaphor}. Each topic is represented as an island and each STA as a tree on that island. We chose this mapping because it makes the abstract STA data tangible and familiar, since people intuitively understand landscapes and vegetation. The spatial proximity of trees then naturally reflects similarity between the STAs. We believe the choice of this metaphor also makes the visualization more playful and engaging.



\begin{figure}[ht]
    \centering
    \includegraphics[width=\linewidth]{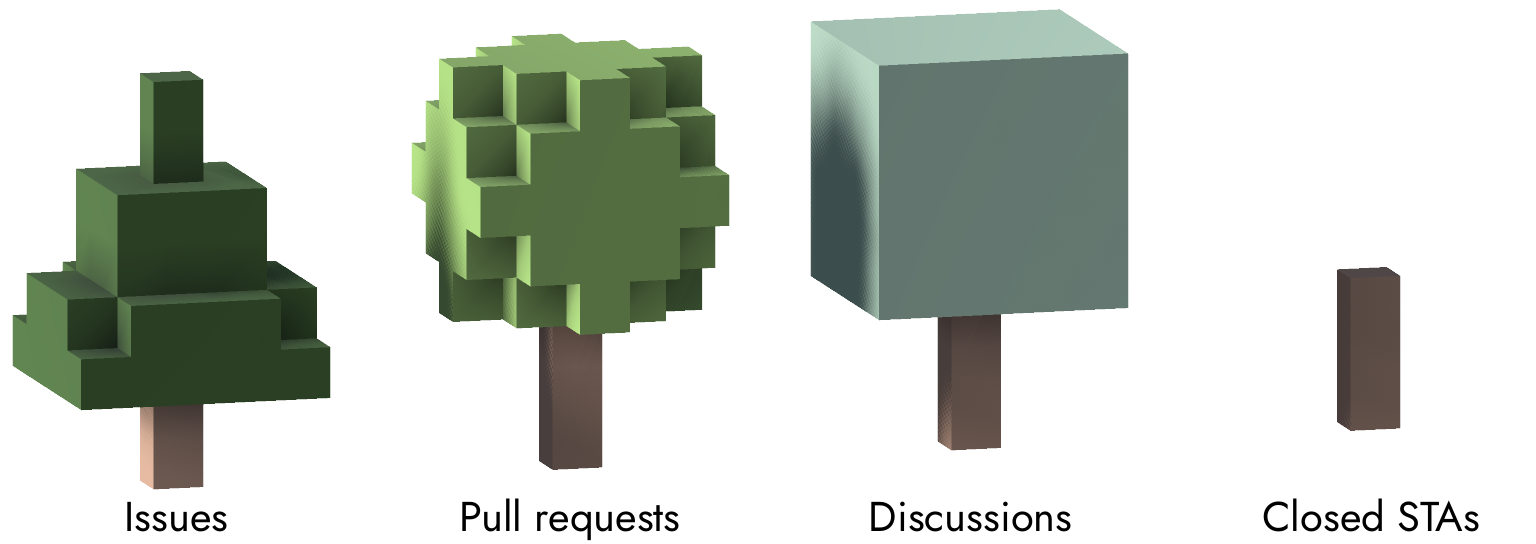}
    \caption{Trees represent types of GitHub socio-technical artifacts. Stubs depict artifacts which have been closed.}
    \label{fig:trees}
\end{figure}

\textbf{Voxel-based glyphs for STAs:} The visualization represents STAs as 3D tree glyphs (\autoref{fig:trees}). Issues are rendered as trees with a conical top, PRs as ball-top trees, and Discussions have a cubical treetop. To improve recognizability, the STA type is also encoded in the treetop color. STAs that have been closed, merged or answered on GitHub can be optionally rendered as stubs, showing that they are no longer considered relevant.

The two-dimensional location of each tree within the world is obtained through a UMAP reduction of its text embedding (\autoref{sec:preprocessing}). Therefore, the distance between trees reflects the semantic similarity of their corresponding STAs. However, tree positions may be slightly adjusted to prevent overlapping.

\textbf{Islands for topics:} Each island is generated from a Delaunay triangulation of the positions of its trees, followed by the removal of triangles that are overly long or overlap with other islands, resulting in one or more polygons. These polygons are used to interpolate vertical positions of the trees, converted to a heightmap, and blurred, creating a smooth terrain.

\textbf{The Third Dimension:} We use the third dimension---island height---to represent STA activity. Specifically, the more comments an STA has, the higher the ground upon which its tree stands. To show how the topic gains or loses traction in time, we calculate island terrains in a sliding window of configurable length. For example, a sliding window of one year means that the terrain shows all STAs with comments not older than one year from the visualized time.

\begin{figure}[ht]
    \centering
    \includegraphics[width=\linewidth]{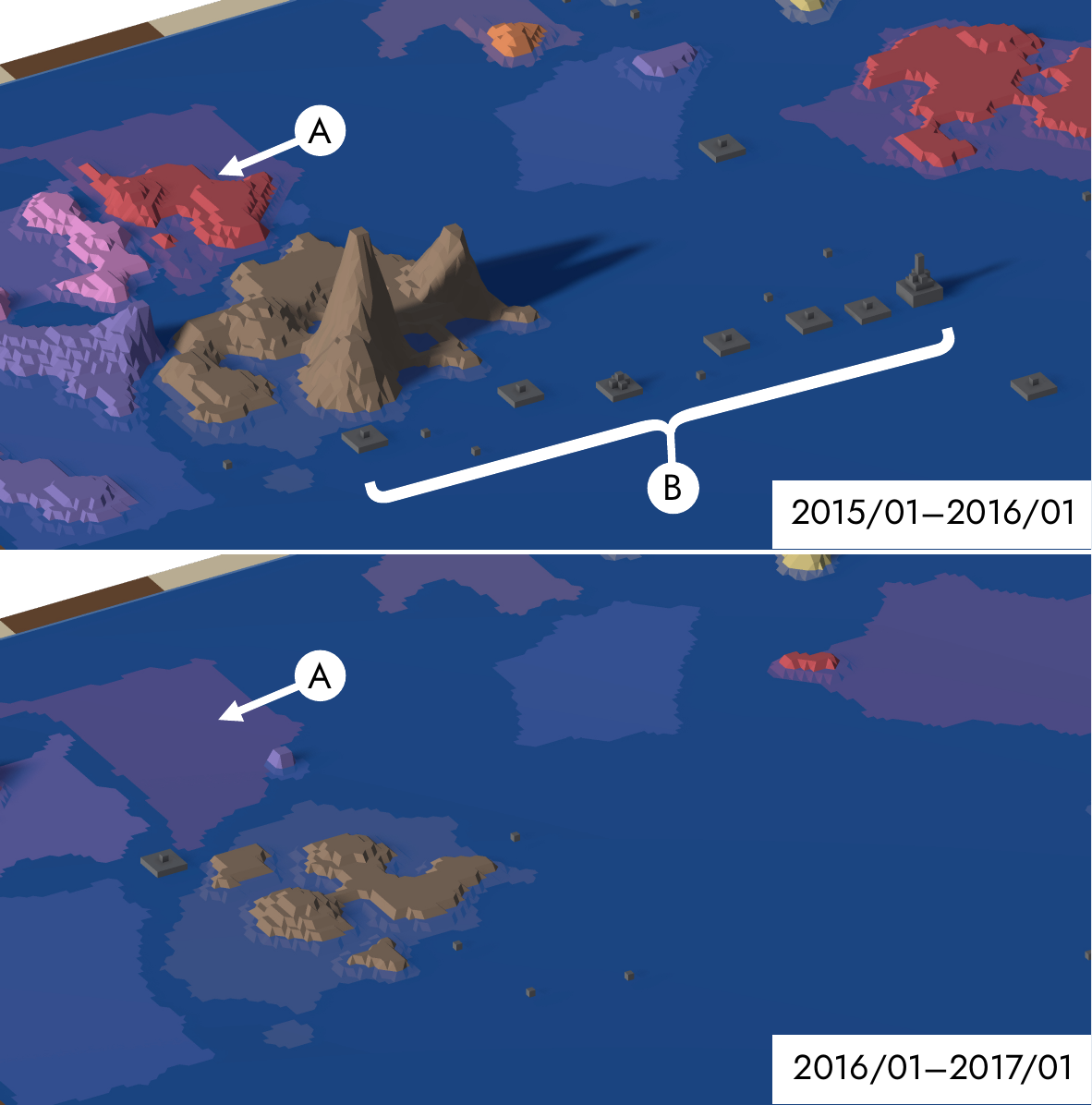}
    \caption{Islands rise (top) and sink (bottom) with the activity of topics in a one-year sliding window~(A). Rocks represent outlier STAs~(B). Terrain is shown without trees for clarity.}
    \label{fig:islands-no-trees}
\end{figure}

The consequent visual effect (\autoref{fig:islands-no-trees}) is that topic islands rise out of the ocean when they are discussed, and sink back in when they are no longer active within the sliding window, thus showing the evolution of the topic's discourse.

The shape of each island is visible just below the ocean's surface even if the topic has no active STAs within the selected sliding window. This informs the user about the proportion of the currently selected timeframe with respect to the project's entire history. Consequently, if the entire evolution of the project is visualized, no submerged islands are rendered. For example, \autoref{fig:islands-no-trees}~\encircle{A} shows a topic that, while of interest until the end of 2015, was never discussed in 2016.


The island color is selected randomly from a palette of eight soft colors, excluding tints of green and blue, reserved for the background (\ie ocean) and the trees, to improve contrast.

Not all STAs belong to a topic. HDBSCAN* may classify some artifacts as outliers. To differentiate them from the topic islands, the outlier trees stand on rocks in the middle of the ocean (\autoref{fig:islands-no-trees}~\encircle{B}). Outlier rocks have a \textit{voxel-based} aesthetic to differentiate them from the {\em low-poly} style of the islands.

\textbf{The t(r)opical islands map:} The 2D area of the visualization map can optionally be proportional to either the size of the project's codebase or the ``editing'' activity on its source code across its history. Specifically, it can correspond to one of the following metrics: The repository's size in kilobytes, its file count, or the number of thousands of lines of code changed across the repository's lifetime (changed LoC).

We compute changed LoC as the cumulative sum of the size of the patch resulting from each Git commit in the history of the repository's default branch. This allows for visual comparison of multiple projects with different sizes and histories and is complemented by the number of STAs generated by its community.

The changed LoC metric is also the most comparable to the number of STAs used to generate the island meshes, since both take the repository's complete history into account. As a result, the ratio of oceans to island-covered areas represents the repository's ratio of code-related to STA-related activities.

Throughout \ritgard's evaluation (Section~\ref{sec:case-studies} and \ref{sec:user-study}), we relied on the changed LoC metric, as it produced the most informative results. We manually looked for outliers in changed LoC (\eg large automatic refactorings, style changes resulting in an artificially large number of modified lines) and we found none, ensuring the accuracy of the representation.


The final visualization is interactive. The user can zoom, pan, and tilt the camera. Each island and tree also gets highlighted with an outline when hovered over with a mouse.

The title of the hovered STA (or topic for islands) is shown in a status bar at the top of the view along with basic statistics. When a tree is clicked while holding Control/Cmd, the corresponding STA is opened in the user's default web browser for further exploration. The visualization can be moved forward or backward in time and the length of the sliding window is configurable.


\section{Case Studies} \label{sec:case-studies}

Leveraging \ritgard, we present several case studies to illustrate the exploration, visualization, and insights elicited by the STA evolutionary analysis of a project. To illustrate our approach, we analyze JetUML, Lume, and Git for Windows.

\autoref{tab:case-studies} shows descriptive statistics of the studied projects. We select Lume and JetUML for the familiarity of some of the authors with them, also as users, and Git for Windows based on the currently supported upper bound on the number of manageable STAs in \ritgard's pipeline.

\begin{table}[ht]
    \SetTblrInner{rowsep=0.0pt, colsep=5.0pt}
    \centering
    \vspace{-1.5mm}
    \caption{Descriptive statistics of the analyzed projects.}
    \label{tab:case-studies}
    \begin{tblr}{
        colspec = {lrrrr},
        row{odd} = {bg=white}, row{even} = {bg=tablelightgray},
        row{1} = {font=\bfseries, fg=white, bg=tabledarkgray},
        stretch=0, rows={ht=\baselineskip}
    }
        \SetCell[]{l} Metric & \SetCell[]{r} Git for Win. & \SetCell[]{r} JetUML & \SetCell[]{r} Lume & \SetCell[]{r} DotVVM \\
        Created on    & 2014-08-22 & 2015-01-07 & 2020-09-07 & 2014-11-05 \\
        Mined on      & 2025-10-21 & 2025-10-21 & 2025-10-21 & 2026-02-10 \\
        History       & 11+ years  & 11+ years  & 5+ years   & 11+ years\\
        Files         & 3,540      & 420        & 562        & 2,560 \\
        Changed LoC   & 31,029,161 & 853,677    & 630,481    & 12,419,651 \\
        Issues        & 4,564      & 420        & 453        & 874 \\
        PRs           & 843        & 151        & 233        & 1,114 \\
        Discussions   & 226        & 4          & 93         & 0 \\
        STAs          & 5,633      & 575        & 779        & 1,988 \\
        Stars         & 8.9k       & 0.7k       & 2.2k       & 0.8k \\
    \end{tblr}
    \vspace{-1.5mm}
\end{table}



\subsection{Lume: STAs in Static Project Snapshots}

Lume\urlfootnote{https://github.com/lumeland/lume} is a static website generator. We analyze a snapshot of the project with a sliding window that takes into account all of its history (shown in \autoref{fig:teaser}). At a glance, Issues are the dominant artifact type, followed by PRs and then Discussions.

Considering map size and the area occupied by the islands, the project has more STAs than code (w.r.t. the number of changed LoC). Both observations are corroborated by \autoref{tab:case-studies}.

Issues are responsible for the tallest peaks of the islands, with the highest having 25 comments. They are distributed across a wide range of topics, roughly corresponding to the project's features, such as support for Markdown, JSX,\footnote{Support for HTML tags in JavaScript commonly used by web frameworks.} URL handling, localization, and Lume's command-line interface.

Most PRs (80.0\%) are concentrated on a single island (\autoref{fig:teaser}~\encircle{A}), depicting \topic{Plugin documentation and refactor} (67 PRs, 13 Issues). Most are minor fixes and maintenance activities related to Lume's documentation and plugins. Much like Issues, the rest of the PRs are spread over Lume's features, with five major islands counting at least 10 PRs each.

Discussions are the least common artifact type in Lume, which is apparent from their rarity in the visualization. They are typically not very complex, judging by their highest peak, representing just five comments as the longest Discussion. The most active topics regard visual rendering improvements of URLs and the internals of Lume's rendering and pagination.


\begin{figure*}[ht]
    \centering
    \includegraphics[width=\textwidth]{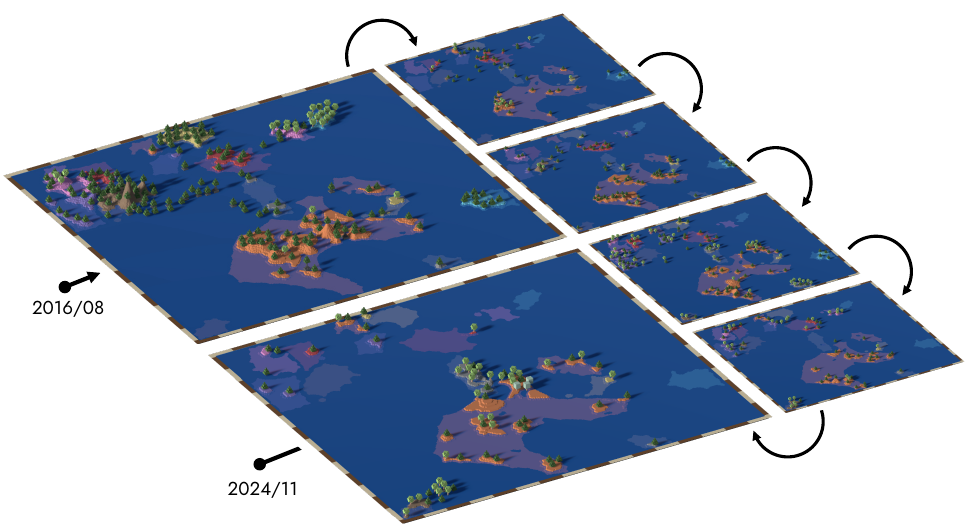}
    \caption{Evolution of JetUML from the perspective of its GitHub STAs. Each arrow represents the passage of 20 months and activity since the previous snapshot. Together, they show 10 years of JetUML's development, since 2015 till the end of 2024.}
    \label{fig:jetuml-evolution}
    \vspace{-2mm}
\end{figure*}

Overall, Lume has a lively community and a good mixture of all three STA types, suggesting that the GitHub repository is a virtual place where developers and users cross paths to discuss technical features and desiderata for the project. This mix is especially visible on the secluded island (\autoref{fig:teaser}~\encircle{B}), where all three STAs are present, representing the \topic{Responsive image plugin} topic and, more in general, image handling.

\insightBox{
    \smallskip
        \boldit{Single snapshot STA visual analysis} allows identifying how the artifacts generated in different communication channels are distributed by topic, making it easy to find the most active conversations and topics.
}

\subsection{JetUML: STAs and Project Evolution}

We chose JetUML,\urlfootnote{https://github.com/prmr/JetUML} a UML (Unified Modeling Language) drawing tool written in Java, as a case study to analyze the evolution of the STAs during the evolution of the project's codebase. We analyze different snapshots of JetUML, taken twenty months apart from each other (\autoref{fig:jetuml-evolution}).



The first snapshot, from August 2016, shows that in the 20 months of the project's existence it has been very active. Issue-based development seems the dominant model for the project, as indicated by the large number of Issues and PRs.

The largest peak is an issue regarding the addition of a copy-paste feature. The largest landmass belongs to the \topic{JavaFX UI and Diagrams} topic, arguably one at the core of the UML sketching tool. Upon closer inspection though, leveraging the hover tooltips with titles of individual PR, most of the PRs are created mainly to do minor refactorings of the codebase and to fix bugs and typos. Issues are quite broadly spread and discussed, but they seldom result in PRs, which are probably reserved for bringing in external contributions from forked repositories not discussed in any specific issue (as suggested by the separate island with almost exclusively PRs).

Summarizing the trend visible in the following snapshots (the four smaller maps in \autoref{fig:jetuml-evolution}), the overall liveliness of the project sees a decline in community participation and involvement. Entire landmasses and their corresponding topics disappear completely under the ocean, and new topics are very few and limited in space. Fewer issues are created, while PR activity remains low but at a steady rate for the years to come. The number of active STAs never reaches again the high mark set in the first snapshot. However, at the end of 2024, the project is still actively developed, with multiple PRs discussed.


\begin{figure*}[ht]
    \centering
    \includegraphics[width=\textwidth]{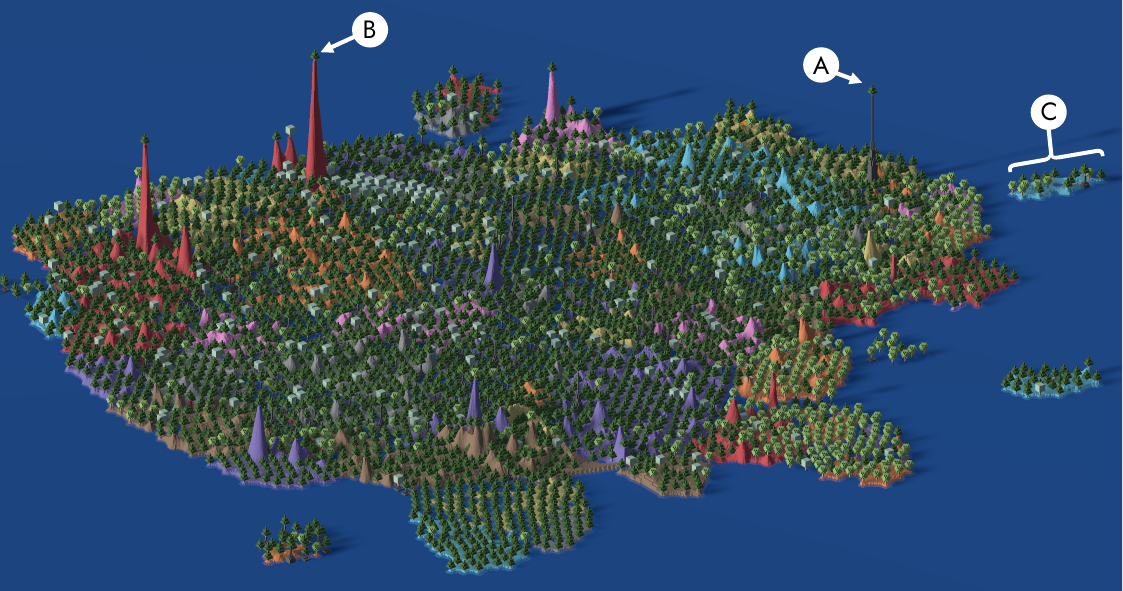}
    \caption{The main landmass of the \textit{Git for Windows} visualization. Peaks of most discussed STAs~(A,~B). Secluded topic of a dependency's updates~(C).
    }
    \label{fig:git-overview}
    \vspace{-2mm}
\end{figure*}


Our observations about the project's evolution suggest that either the project's initial excitement died down except for its core audience, or the initial onslaught of issues got resolved after the first snapshot. Notable is also the complete absence of GitHub Discussions everywhere except for the last snapshot, a clearly late and ineffective attempt to use the new GitHub feature (publicly released in December 2020~\cite{hata_2021}).

Another reflection elicited by the partially disappearing landmass of the central island is that the underwater parts of the islands also communicate information about the activity of topics being discussed in the past or to be discussed in the future. We experimented with keeping stubs of the closed STAs visible underwater, to differentiate between past and future, but the visualization was too cluttered to be informative. Future work in this regard is needed to understand how to represent closed STAs to be more revealing at a glance, without animations, and with minimal clutter.

\begin{figure}[ht]
    \centering
    \includegraphics[width=\linewidth]{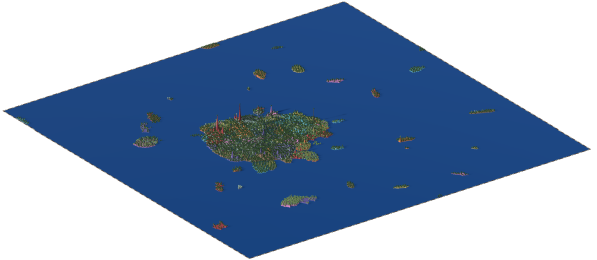}
    \caption{The mostly oceanic \textit{Git for Windows} map with the main landmass of topics surrounded by smaller, scattered islands.}
    \label{fig:git-minimap}
    \Description{A small image shows the whole visualization of Git for Windows. It is mostly oceanic. Close to the center is the large landmass from the previous figure. There are also many small islands floating in the ocean.}
\end{figure}

\begin{figure*}[ht]
    \centering
    \includegraphics[width=\textwidth]{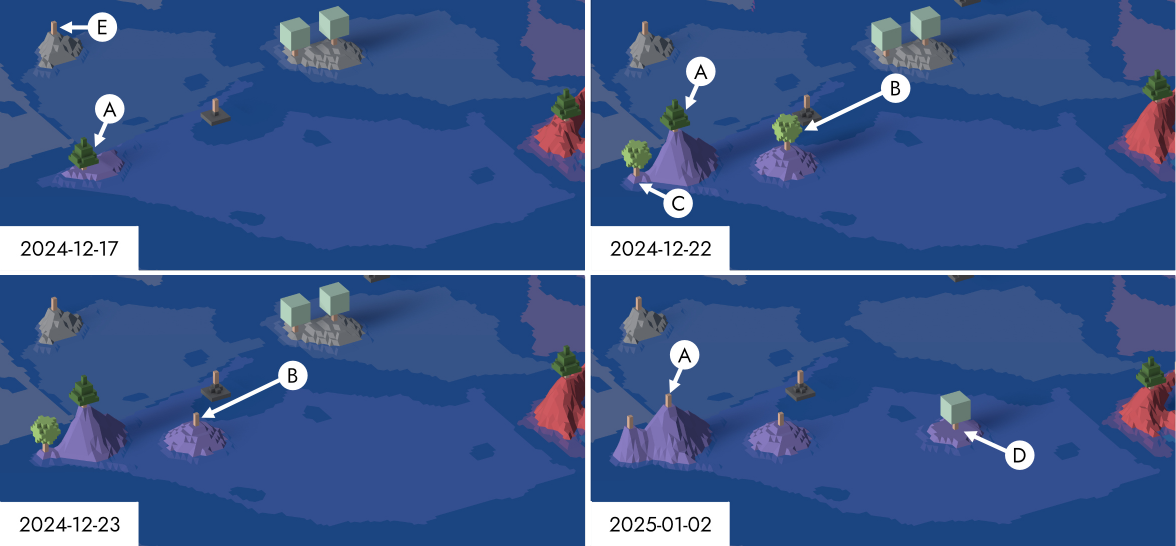}
    \caption{Example of interconnected STAs (four snapshots, one-month sliding window). Following a bug report (A) regarding the \textit{credential-cache} there is a first attempt at an unmerged PR (B), followed by a successful one (C). Discussion (D) appears one week after, about a related topic. A thematically close yet unrelated STA (E) about \topic{Git authentication issues} is already closed.}
    \label{fig:git-credential-cache}
\end{figure*}

\insightBox{
    \smallskip
	\boldit{STA evolutionary analysis} allows comparing the activity of different artifacts in different phases of a project. More importantly, it shows by contrast how the topics evolve across time, giving insights on the project's core aspects, attention and discussion elements important to the users.
}

\subsection{Git for Windows: STAs in the Large} \label{sec:git-for-windows}

To test the efficacy of our approach on a large project with a long history, we analyze Git for Windows\urlfootnote{https://github.com/git-for-windows/git} (GfW), an actively developed Windows-specific fork of Git. This project differs from the other case studies in the size of its codebase, with more than 3.5k files in the current version, in the length of its evolution, with more than 31M changed LoC, and in the number of STAs generated in the 11 years of history: one order of magnitude larger than the previously analyzed projects. In \autoref{fig:git-overview}, we show the main landmass of the topic islands of GfW and its STA trees, as of October 2025.

Being a fork has a notable consequence on the visualization. There is a lot more ocean than there is landmass (\autoref{fig:git-minimap}), because issues related to the upstream Git codebase are discussed over a mailing list rather than a GitHub repository.

The most noticeable features of the visualization's ``mainland'' are the tall peaks. These correspond to the most active STAs across the lifetime of GfW, marking some of its more significant events. For instance, \encircle{A} is an Issue about non-inclusive naming,\urlfootnote{https://github.com/git-for-windows/git/issues/2674} such as using \texttt{master} as the name of the default branch. This is a very polarizing topic with a vocal part of the community involved in the request for changes. Another example is about support for the ARM64\urlfootnote{https://github.com/git-for-windows/git/issues/3107} architecture \encircle{B}, a technical improvement requiring careful consideration and planning, but also potentially bringing new users to the project. As outliers, both of these issues have over 150 comments, so to prevent the peaks from being overly tall, the heights of the islands are normalized to a configurable maximum.

The smaller islands away from the mainland represent self-contained topics. For example, the topic \topic{PCRE2 version update} \encircle{C} concerns updates of GfW's dependency handling regular expressions. There is also the \topic{Git for Windows releases} topic, which contains most of the project's GitHub Discussions. For every release, a new Discussion is automatically opened and users can comment on it (\eg report problems with the specific release).

The visualization shows that some topics are persistently active over the project's history. These topics tend to be related to Windows-specific behavior (\eg case-insensitive paths) and advanced features (\eg submodule recursion). Other topics, such as \topic{Missing .mo localization}, re-emerge repeatedly with long periods of silence in between.


Despite the size of the GfW project, the visualization tells small developer stories as well. \autoref{fig:git-credential-cache} depicts what occurred in December 2024, regarding \topic{Git Credential Manager updates}. First, Issue~\encircle{A}\urlfootnote{https://github.com/git-for-windows/git/issues/5314} was created to report a problem with the \texttt{credential-cache}. Several days later, the Issue was addressed by PR~\encircle{B}. This PR has been closed without being merged and was then superseded by PR~\encircle{C}. This later PR was merged successfully without much discussion, as shown in the visualization. Ten days later, the Issue and both PRs are closed. However, a Discussion~\encircle{D}\urlfootnote{https://github.com/git-for-windows/git/discussions/5338} is created, asking about an error with a component related to \texttt{credential-cache}.


Finally, clustering limitations become more pronounced in large repositories, such as GfW, where many topics ``fight for space on the map''. As a result, not all close STAs are related. For instance, \autoref{fig:git-credential-cache} shows an Issue~\encircle{E},\urlfootnote{https://github.com/git-for-windows/git/issues/5306} unrelated to \texttt{credential-cache}. It is, in fact, part of a different topic (\topic{Git authentication issues}) which is, however, thematically close to \texttt{credential-cache}, explaining its positioning.

The case of GfW shows how the visualization provides insights in four directions. From a high-level overview to a detailed analysis of short time spans, it provides information about the STAs, with insights into the development and management processes that generated them. Another axis goes from a static to an evolutionary perspective. Static visualizations show conversation hotspots as peaks, while analyses taking into account the temporal dimension show long-lived topics, recurring ones, and short single-concern stories.

Considering that the Discussion~\encircle{D} from \autoref{fig:git-credential-cache} remains unanswered (as of October 2025) the visualization can also be useful to pinpoint potentially overlooked STAs. Finally, this case study shows that, although improvements to the clustering step would benefit the visualization, even embedding only the STA titles and labels is sufficient to obtain a useful clustering to extract meaningful insights in real world scenarios.

\insightBox{
    \smallskip
	\boldit{STA analysis scales} to large long-lived repositories and provides insights in the management and development processes, with emphasis on the participation and engagement levels of different communities.
    \smallskip
}


\vspace{-2mm}

\section{User Study} \label{sec:user-study}

We conducted a user study with 34 participants (31M/3F, age 21--43, mean $M\!=\!26.53$, $\mathit{SD}\!=\!4.49$) recruited through personal contacts and snowball sampling~\cite{goodman_1961}, including CS students, software engineers, project managers, and academics. Sessions were conducted in-person (30) and remotely (4).

The study's main goal was to assess the interpretability of the visualization and, thus, the viability of our approach. The study design mirrored the case studies in \autoref{sec:case-studies}, focusing on snapshot analysis, evolutionary analysis, and large-scale visualization in three sets of tasks. The Lume and JetUML projects were used for the first two sets. However, we used \textit{DotVVM}\urlfootnote{https://github.com/riganti/dotvvm} (\autoref{tab:case-studies}), a smaller example compared to Git for Windows, for the third set of tasks to mitigate our concern that participants would focus on the prototype's performance limitations rather than the visualization approach itself.

We formulated five hypotheses:
\begin{enumerate}[beginpenalty=10000]
    \item[\textbf{H1}] the topic islands metaphor provides a readable representation of GitHub conversations (STAs) and their topics; 
    \item[\textbf{H2}] the visualization enables users to derive insights about repository activity and topics;
    \item[\textbf{H3}] the playful visual design does not negatively affect readability; 
    \item[\textbf{H4}] the visualization is perceived as useful for software engineering tasks; and 
    \item[\textbf{H5}] the visualization remains readable for projects of different sizes.
\end{enumerate}

After providing consent for the study, participants completed an initial questionnaire about their demographics, background, current roles, and GitHub experience. We presented the use of \ritgard in a video tutorial. The three sets of tasks (10 tasks in total) were designed to explore and use all of \ritgard's features, while thinking aloud. The correct answers for six analytical questions were determined by manually inspecting the repositories. Four remaining questions were exploratory and open-ended. The participants then completed a post-task questionnaire evaluating the visualization. The average session duration was 56:42 minutes ($\mathit{SD} = \textrm{19:25} \textrm{ min}$).

\paragraph*{Interpretability and Insights}

\begin{figure}[t]
    \centering
    \includegraphics[width=\linewidth]{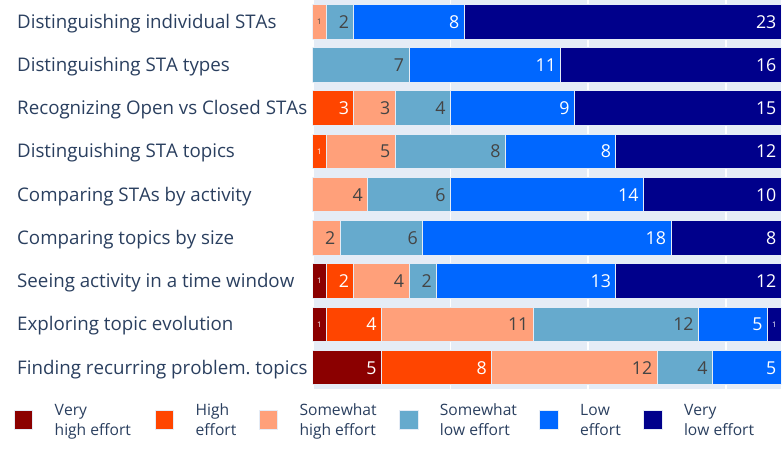}
    \caption{Reported effort for activities in \ritgard.}
    \label{fig:us-effort}
\end{figure}

Basic activities such as distinguishing STAs or comparing topics by size were considered \textit{Somewhat low effort} or easier by most participants ($>80\%$, \autoref{fig:us-effort}). Other questionnaire responses confirm this observation: Interpretation effort was very low (identification tasks \mbox{$M=1.89$}, comparison tasks \mbox{$M=2.09$} on a six-point Likert scale), both significantly below the midpoint (Wilcoxon $p\!<\!0.001$, effect size $r\!\approx\!0.85$--$0.88$), indicating that users could easily read the visualization, supporting \textbf{H1}.

Activities involving evolutionary analysis were perceived as more demanding because participants often had to repeatedly step through the project's history. Several participants suggested that a timeline could improve navigation, for example:

\begin{displayquote}[\textbf{P18}]
I feel like a timeline in the UI would help me immensely with understanding what timeframe I'm looking at and how quickly I'm moving through it. Without anything like that, I felt lost in what the current step and sliding window mean.
\end{displayquote}

Consistent with this observation, participants reported moderate effort when extracting insights (\mbox{$M=3.27$}, Wilcoxon \mbox{$p=0.031$}, \mbox{$r=0.32$}). Despite this perceived effort, analytical questions were answered with high accuracy (mean: 89\%, range: 76--97\%), significantly above chance (binomial tests \mbox{$p\le0.002$}). This indicates that users were able to derive correct insights from the visualization, supporting \textbf{H2}.

\paragraph*{Playfulness}

\begin{figure}[t]
    \centering
    \includegraphics[width=\linewidth]{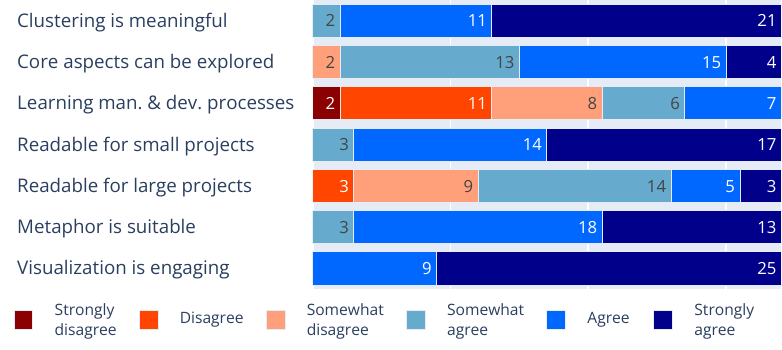}
    \caption{Agreement with statements about \ritgard.}
    \label{fig:us-agreement}
\end{figure}

Participants generally evaluated the island metaphor positively, as shown in \autoref{fig:us-agreement}. 22 respondents (65\%) agreed that the visualization balances playfulness and readability well, while only one preferred a more abstract visualization. One participant summarized this enthusiasm:

\begin{displayquote}[\textbf{P16}]
Give me a diorama, a hologram on my table, so that I always see how the project is doing.
\end{displayquote}

Questionnaire responses support this perception: The playful design was rated very positively (\mbox{$M=5.52$}, Wilcoxon \mbox{$p<0.001$}). Moreover, perceived playfulness did not correlate with interpretation effort (Spearman \mbox{$\rho=-0.25$}, \mbox{$p=0.16$}), indicating that the playful design does not negatively affect readability and supporting \textbf{H3}.

\paragraph*{Usefulness}

\begin{figure}[ht]
    \centering
    \includegraphics[width=\linewidth]{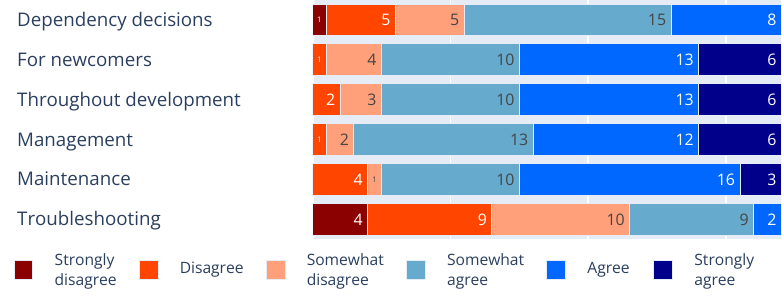}
    \caption{Perceived suitability of \ritgard for different use cases.}
    \label{fig:us-usecases}
\end{figure}

Participants considered \ritgard particularly useful for onboarding newcomers, monitoring development progress, project management, and maintenance (\autoref{fig:us-usecases}). It was also viewed as somewhat useful for evaluating potential dependencies, although several noted that repository statistics might suffice for this task. Conversely, the visualization was generally considered unsuitable for troubleshooting, though some suggested that full-text search could improve this scenario. Usefulness ratings were clearly above neutral (\mbox{$M=4.11$}, Wilcoxon \mbox{$p<0.001$}, \mbox{$r=0.76$}), supporting \textbf{H4}.

\paragraph*{Scalability}
Participants generally found the clustering meaningful. Readability was rated high for smaller repositories such as Lume and JetUML but decreased for larger projects like DotVVM because individual trees became smaller and harder to distinguish. Some participants also attributed this to slower interaction response times in the prototype.

These observations are reflected in the questionnaire responses: Readability ratings were significantly above neutral for both small (\mbox{$M=5.41$}, Wilcoxon \mbox{$p<0.001$}, \mbox{$r=0.89$}) and large repositories (\mbox{$M=3.88$}, Wilcoxon \mbox{$p=0.028$}, \mbox{$r=0.33$}). However, a paired Wilcoxon test showed that large projects were rated significantly harder to read (\mbox{$p<0.001$}), suggesting that while the visualization remains usable at larger scales, its readability decreases as project size increases, partially supporting \textbf{H5}.

Finally, when asked to compare \ritgard with GitHub's user interface, 23 participants (68\%) considered the two complementary, emphasizing that \ritgard provides a high-level overview of project activity that traditional interfaces lack.

Overall, the results provide evidence supporting the interpretability, perceived usefulness, and suitability of the topic islands metaphor, supported and enriched by an engaging visualization. Due to space constraints, the results presented here are condensed; the full tasks, questionnaire, responses, and analyses are included in the \hyperref[replipackage]{replication package}.


\section{Discussion} \label{sec:discussion}

The case studies and user study demonstrate that our visualization approach, as implemented in the \ritgard prototype tool, allows for a comprehensive analysis of a software project from the perspective of its GitHub STAs. It depicts the project's main features, problematic areas, and even its team's development habits. STA analysis with \ritgard is a step towards a better understanding of how the developer and user attention shifts during the project's lifetime, across different features and concerns. \ritgard achieves this while also balancing its readability and playfulness.

\subsection{Lessons Learned}

We present the main reflections about our approach that emerged during its prototype implementation, and our validation through the user study.

\textbf{The target audience is diverse:} The visualization can be useful for many stakeholders. Software developers can use it to ``chart'' the topics their community is interested in, look for problems they encounter most frequently, or to find STAs that might otherwise be overlooked. For newcomers to a project, a clear high-level visualization of the STAs can impact their decision to rely on it as a dependency. The project's core strengths, pain points, and a measurement of its development activity emerge as clear aspects from such analysis. The approach can also be beneficial for existing users in bringing them up to speed with the project's latest development activity, with an appropriate use of the sliding window. Finally, for project managers, the visualization can serve as an intuitive overview of project status and development velocity.

\textbf{Garbage in, garbage out:} The resulting visualization is only as good as the clustering of the STAs. Bad clustering with interleaved and scattered points results in a landscape that is difficult to read. To ensure the best clustering possible, we pre-processed the STAs removing elements that would add noise with wrong ``artificial'' similarities (\eg links, code snippets, markdown syntax); we embedded the artifacts using the best available embedding model for semantic textual similarity; and we devised an island generation procedure to produce smooth surfaces and reduce visual anomalies and artifacts.

\textbf{One size does not fit all:} Although all visualizations in the case studies and the user study rely on the number of changed LoC as a measure for the size of the map, this is not suitable for all projects. Notably, there are GitHub repositories serving exclusively as issue or proposal trackers, containing very little code compared to an enormous number of STAs. For this kind of projects, the map metric can be turned off altogether and the visualization's size is automatically set to minimize tree overlaps. With a growing size of the project, the trees also become hard to distinguish, as participants of the user study pointed out. Therefore, for large projects, aggregation should be performed on multiple levels, not just the islands.

\textbf{Verdant islands of healthy discourse:} The islands can be used to gauge the ``health'' of a topic. If an island is covered in trees, it may be an outstanding problematic area. If it also has a mixture of STA types, it might mean that the topic is important for users and developers alike. However, this metaphor has limitations. For example, it is not clear whether an island is submerged because it is either no longer active or not active yet. Similarly, a tree stub indicates that the STA is closed but not the reason for its closure. For example, PRs can be either closed and successfully merged or closed and discarded. Consequently, a submerged or barren island does not imply that the topic is ``unhealthy'', but only that it is inactive or its STAs are closed, respectively. We plan to build upon the topic islands metaphor in a future extension to better represent not just the liveliness of a topic but also its health status.

\subsection{Limitations and Future Work} \label{sec:limitations}

Although the approach has clear advantages and use cases, there remain some conceptual and technical limitations.

The end result depends on the clustering of the STAs, which in turn depends on the embedding model, which may be computationally demanding. To partially mitigate this, we decoupled the data miner, the processing steps, and the visualizer, and let them communicate through a shared exchange format in JavaScript Object Notation (JSON). This allows offloading the text embedding step to a separate machine with a GPU suitable for the chosen model. The hardware requirements can also be relaxed by selecting a less demanding model, which would make the approach accessible for a broader range of users at the cost of less accurate clustering and topic names.

The pre-processing is also a limiting factor in the quality of the clustering. For example, Issue and PR templates, commonly used to standardize the format of bug reports and other common STAs, are currently not taken into consideration, thus potentially negatively affecting the text embeddings. This may result in large islands consisting mostly of STAs of one ``type'' connected to a specific template. Some STAs can also be too large to be embedded at once, due to limited context size of the embedding model or its demands on GPU memory. For example, in the GfW case study, we excluded STA bodies and comments from the embeddings, due to memory constraints. This shows that it is still possible to obtain meaningful insights with our visualization despite the trade-offs with the embeddings. In the future, we plan to conduct experiments by appropriately chunking the STAs to keep more context and re-combining them through the embeddings.

The data miner has its specific limitations. For GitHub Discussions, it retrieves only the main comments without their inner replies (\ie the threads), which might lead to the underrepresentation and mis-classification of Discussions with respect to other STAs. Moreover, for all closed STAs, we mine only the \textit{last} timestamp of their closure. Therefore, the visualization ignores STA re-openings.

Our design choices to make the visualization engaging and playful resulted in a visually appealing, readable, and informative visualization, but they also come with trade-offs. For example, when the number and density of the visualized STAs is high, the resulting tree foliage can obstruct the view of the underlying terrain. We partially mitigate this shortcoming by including a UI option to hide the trees when the individual STAs are not of primary importance.

\ritgard is a proof-of-concept research prototype. Rendering performance was not prioritized throughout its development. Consequently, it is limited in the number of STAs it can handle. Concrete upper bounds depend on the available hardware. In the future, if performance becomes a focus, we can leverage typical rendering trade-off ``tricks'' to scale up the number of analyzable projects, also aiming towards real-time animation of the evolution. So far, we focused on experimenting with the core visualization. In the future, the most significant improvement will be a proper timeline, intuitive sliding window controls, and better label placement as suggested by the study's participants.

The user study may not be fully representative, as many of the participants were undergraduate or graduate students (41\%), few had experience with project management (6\%), and most were male (91\%).

The tasks featured in the user study were chosen to evaluate the interpretability of the visualization; therefore, the reported usefulness is based on the participants' perceptions. Similarly, although the participants were asked to compare \ritgard to GitHub's native UI, these observations are not a substitute for a direct task-based comparison.

Despite the phrasing of the questions and explicit instruction to focus on the visualization, participants sometimes did not distinguish between the visualization and the prototype. This skews the results toward negative answers due to assessing the prototype's performance and user experience, rather than the visualization itself.

\pagebreak

None of the limitations discussed here are unexpected, and they provide avenues for future work. \ritgard can be extended to mine other platforms (\eg GitLab, Bitbucket, Gitea). The tree glyphs can be made more expressive to depict additional information about the represented STAs. Lastly, the visualization could take the source code of the system into account, connecting the STAs to the related source code actions, opening new possibilities to expand the metaphor.


\section{Related Work} \label{sec:related-work}

GitHub and its STAs are an active research subject. Gousios \etal conducted a foundational study on millions of PRs to learn how developers use the pull-based development model~\cite{gousios_2014}. Hata \etal researched GitHub Discussions specifically, showing their worth for spawning new Issues and PRs~\cite{hata_2021}. Siddiq \etal used machine learning (\ie a BERT-based model) to automatically classify Issues and assign them labels~\cite{siddiq_2022}. Venigalla \etal showed the documentation potential of various STAs and studied their topics~\cite{venigalla_2024}. Barbosa \etal focused on the social dimension of GitHub STAs, conducting a study into how factors such as communication dynamics and team size impact code quality in PRs~\cite{barbosa_2023}.

Nevertheless, GitHub is not the only platform of interest. For example, Barua \etal used latent Dirichlet allocation to extract topics and trends form StackOverflow posts~\cite{barua_2014}. Parra \etal compared the Gitter and Slack messaging platforms, produced a manually annotated dataset of Gitter messages, and evaluated several ML techniques for their automatic labeling. Warrick \etal studied developer ecosystems (Python, Go, and Node.js) as socio-technical systems, mined their mailing lists, producing the OCEAN dataset~\cite{warrick_2022}. All of these examples showcase the diversity of research into software-related socio-technical artifacts.

Topic modeling lies at the core of our visualization approach. It is a set of techniques used to extract latent topics from a body of text in a natural language. These techniques have been developed for decades and in recent years benefited from the use of neural networks. We refer to two surveys by Wu \etal~\cite{wu_2024} and Zhao \etal~\cite{zhao_2021}, respectively, for an overview of topic modeling and its use of neural networks.

Landscapes and islands have been used to aid in program comprehension before. Kuhn \etal used latent semantic indexing and multidimensional scaling to produce a 2D thematic map of a codebase~\cite{kuhn_2008}. Atzberger \etal took a similar approach with Latent Dirichlet Allocation in their Software Forest, where each software entity becomes a tree~\cite{atzberger_2021}. Misiak \etal visualized packages and their Java classes as islands and buildings both in 3D and virtual reality~\cite{misiak_2018}. While the aforementioned works use visual metaphor similar to ours, they visualize code artifacts, not STAs.

Besides the island metaphor, other approaches include that of Fiechter \etal, representing GitHub Issues through their \textit{issue tales}, offering multiple 2D issue visualizations~\cite{fiechter_2021}. However, their approach focuses mostly on the lifecycle of issues, whereas \ritgard's is about providing an evolutionary overview of the whole repository with multiple types of STAs.

Our design choices to make the visualization more engaging, such as representing the STAs with tree glyphs, make our approach related to gamification, which is a lively topic in software engineering in both theory and practice.

For example, Dal Sasso \etal, investigated the usage of game elements and presented a conceptual framework with guidelines on which elements to consider for software engineering tasks~\cite{dal_sasso_2017}. Another example is a gamified visualization designed by Heidrich \etal to foster program comprehension, taking into account developer requirements and getting positive feedback from their players~\cite{heidrich_2025}.

Gamification is connected to the concept of \textit{playfulness} in adults---inclination towards creativity, curiosity, sense of humor, and related traits---which has been linked with the ability to meet problems with an open mind and accept failure, increasing work performance~\cite{guitard_2005}. Playfulness can be a core characteristic of visualization. For example, Schindler \etal employed it when visualizing biological models, producing paper-based ``physicalizations''~\cite{schindler_2022}.



\section{Conclusion} \label{sec:conclusion}

We presented a visualization approach for the analysis of socio-technical artifacts in software projects hosted on GitHub. The visualization depicts each topic extracted from the artifacts as an island in an ocean, the area of which represents the size of the code repository. As topics become more or less discussed, the islands rise in and out of the ocean. The topics themselves are extracted by converting each STA into a vector embedding, clustering these vectors, extracting keywords, and feeding them into a large language model to produce a human-readable label for the island.

We provide insights into the analyzed projects from static and evolutionary perspectives, at multiple levels of granularity. The \ritgard prototype requires optimization and user experience improvements to be usable in practice. However, as shown by the user study, it can support software developers to learn about a project, open-source software maintainers in managing their communities, and it can also serve users of these projects to uncover problematic areas discussed by different stakeholders in different communication channels.

\section*{Replication package}\label{replipackage}

To allow for our work to be verified and replicated, the case study datasets, the user study materials, the participant responses, the topic naming prompt template, the source code, a tutorial video, and builds of \ritgard are available at \break \replipackage

\section*{Acknowledgements}

We thank the participants of our user study and our colleagues who helped refine the study before it was conducted. Computational resources were provided by the e-INFRA CZ project (ID:90254), supported by the Ministry of Education, Youth and Sports of the Czech Republic. We also gratefully acknowledge the financial support of the Swiss National Science Foundation (SNSF) through the project ``FORCE'' (SNF Project No. 232141).

\pagebreak
\bibliographystyle{IEEEtran}
\bibliography{IEEEabrv,references.bib}

@inproceedings{alkadhi_2017,
  title     = {Rationale in Development Chat Messages: {An} Exploratory Study},
  note      = {\doi{10.1109/MSR.2017.43}},
  booktitle = {International Conference on Mining Software Repositories ({MSR})},
  author    = {Alkadhi, Rana and Lata, Teodora and Guzmany, Emitza and Bruegge, Bernd},
  year      = {2017},
  pages     = {436--446},
  publisher = {IEEE}
}

@inproceedings{atzberger_2021,
  title     = {Software {Forest}: {A} Visualization of Semantic Similarities in Source Code using a Tree Metaphor},
  note      = {\doi{10.5220/0010267601120122}},
  booktitle = {International Joint Conference on Computer Vision, Imaging and Computer Graphics Theory and Applications ({VISIGRAPP})},
  publisher = {SCITEPRESS},
  author    = {Atzberger, Daniel and Cech, Tim and De La Haye, Merlin and Söchting, Maximilian and Scheibel, Willy and Limberger, Daniel and Döllner, Jürgen},
  year      = {2021},
  pages     = {112--122}
}

@inproceedings{barbosa_2023,
  title     = {Beyond the Code: {Investigating} the Effects of Pull Request Conversations on Design Decay},
  note      = {\doi{10.1109/ESEM56168.2023.10304805}},
  booktitle = {International Symposium on Empirical Software Engineering and Measurement ({ESEM})},
  author    = {Barbosa, Caio and Uchôa, Anderson and Coutinho, Daniel and Assunção, Wesley Klewerton Guez and Oliveira, Anderson and Garcia, Alessandro and Fonseca, Baldoino and Rabelo, Matheus and Coelho, José Eric and Carvalho, Eryka and Santos, Henrique},
  year      = {2023},
  pages     = {1--12},
  publisher = {IEEE}
}

@article{barua_2014,
  title   = {What Are Developers Talking About? {An} Analysis of Topics and Trends in {Stack} {Overflow}},
  volume  = {19},
  note    = {\doi{10.1007/s10664-012-9231-y}},
  number  = {3},
  journal = {Empirical Software Engineering},
  author  = {Barua, Anton and Thomas, Stephen W. and Hassan, Ahmed E.},
  year    = {2014},
  pages   = {619--654}
}

@incollection{bijker_1987,
  title     = {General Introduction},
  author    = {Bijker, Wiebe E. and Hughes, Thomas P. and Pinch, Trevor J.},
  isbn      = {978-0-262-02262-0},
  booktitle = {The Social Construction of Technological Systems: {New} Directions in the Sociology and History of Technology},
  publisher = {MIT Press},
  year      = {1987}
}

@inproceedings{bouraffa_2025,
  title     = {How do Developers Use Code Suggestions in Pull Request Reviews?},
  note      = {\doi{10.1109/CHASE66643.2025.00033}},
  booktitle = {International Conference on Cooperative and Human Aspects of Software Engineering ({CHASE})},
  author    = {Bouraffa, Abir and Pham, Yen Dieu and Maalej, Walid},
  year      = {2025},
  pages     = {227--238},
  publisher = {IEEE}
}

@inproceedings{dal_sasso_2017,
  title     = {How to Gamify Software Engineering},
  note      = {\doi{10.1109/SANER.2017.7884627}},
  booktitle = {International Conference on Software Analysis, Evolution and Reengineering ({SANER})},
  author    = {Dal Sasso, Tommaso and Mocci, Andrea and Lanza, Michele and Mastrodicasa, Ebrisa},
  year      = {2017},
  pages     = {261--271},
  publisher = {IEEE}
}

@article{drechsler_2015,
  title   = {Designing to Inform: {Toward} Conceptualizing Practitioner Audiences for Socio-Technical Artifacts in Design Science Research in the Information Systems Discipline},
  volume  = {18},
  journal = {Informing Science: {The} International Journal of an Emerging Transdiscipline},
  author  = {Drechsler, Andreas},
  year    = {2015},
  pages   = {31--47},
  note    = {\doi{10.28945/2288}}
}

@inproceedings{fiechter_2021,
  title     = {Visualizing {GitHub} {Issues}},
  note      = {\doi{10.1109/VISSOFT52517.2021.00030}},
  booktitle = {Working Conference on Software Visualization ({VISSOFT})},
  author    = {Fiechter, Aron and Minelli, Roberto and Nagy, Csaba and Lanza, Michele},
  year      = {2021},
  pages     = {155--159},
  publisher = {IEEE}
}

@misc{github_2026a,
  title        = {About Issues},
  url          = {https://docs.github.com/en/issues/tracking-your-work-with-issues/learning-about-issues/about-issues},
  journal      = {GitHub Docs},
  howpublished = {GitHub Docs},
  author       = {{GitHub}},
  year         = {2026}
}

@misc{github_2026b,
  title        = {About Discussions},
  url          = {https://docs.github.com/en/discussions/collaborating-with-your-community-using-discussions/about-discussions},
  journal      = {GitHub Docs},
  howpublished = {GitHub Docs},
  author       = {{GitHub}},
  year         = {2026}
}

@article{goodman_1961,
  title     = {Snowball Sampling},
  volume    = {32},
  doi       = {10.1214/aoms/1177705148},
  journal   = {The Annals of Mathematical Statistics},
  author    = {Goodman, Leo A.},
  year      = {1961},
  publisher = {Institute of Mathematical Statistics},
  pages     = {148--170}
}

@inproceedings{gousios_2014,
  title     = {An exploratory study of the pull-based software development model},
  note      = {\doi{10.1145/2568225.2568260}},
  booktitle = {International Conference on Software Engineering ({ICSE})},
  publisher = {ACM},
  author    = {Gousios, Georgios and Pinzger, Martin and Deursen, Arie van},
  year      = {2014},
  pages     = {345--355}
}

@article{gregor_2013,
  title     = {Positioning and Presenting Design Science Research for Maximum Impact},
  volume    = {37},
  note      = {\doi{10.25300/MISQ/2013/37.2.01}},
  journal   = {Management Information Systems Quarterly},
  author    = {Gregor, Shirley and Hevner, Alan R.},
  year      = {2013},
  publisher = {MIS Quarterly},
  pages     = {337--355}
}

@preprint{grootendorst_2022,
  title         = {{BERTopic}: {Neural} Topic Modeling with a Class-Based {TF}-{IDF} Procedure},
  note          = {\doi{10.48550/arXiv.2203.05794}},
  eprint        = {2203.05794},
  publisher     = {arXiv},
  author        = {Grootendorst, Maarten},
  year          = {2022},
  pages         = {1--10},
  archiveprefix = {arXiv}
}

@article{guitard_2005,
  title     = {Toward a Better Understanding of Playfulness in Adults},
  volume    = {25},
  note      = {\doi{10.1177/153944920502500103}},
  journal   = {Occupational Therapy Journal of Research ({OTJR})},
  author    = {Guitard, Paulette and Ferland, Francine and Dutil, {\'E}lisabeth},
  year      = {2005},
  publisher = {Sage},
  pages     = {9--22}
}

@article{hata_2021,
  title   = {{GitHub} {Discussions}: {An} Exploratory Study of Early Adoption},
  volume  = {27},
  note    = {\doi{10.1007/s10664-021-10058-6}},
  number  = {1},
  journal = {Empirical Software Engineering},
  author  = {Hata, Hideaki and Novielli, Nicole and Baltes, Sebastian and Kula, Raula Gaikovina and Treude, Christoph},
  year    = {2021},
  pages   = {1--32}
}

@inproceedings{heidrich_2025,
  title     = {Towards Higher Motivation to Perform Software Comprehension Tasks Through Gamification},
  note      = {\doi{10.1109/VISSOFT67405.2025.00018}},
  booktitle = {Working Conference on Software Visualization ({VISSOFT})},
  author    = {Heidrich, David and Gökmen, René and Schreiber, Andreas and Bichlmeier, Christoph},
  year      = {2025},
  pages     = {79--83},
  publisher = {IEEE}
}

@inproceedings{kuhn_2008,
  title     = {Consistent Layout for Thematic Software Maps},
  note      = {\doi{10.1109/WCRE.2008.45}},
  booktitle = {Working Conference on Reverse Engineering},
  author    = {Kuhn, Adrian and Loretan, Peter and Nierstrasz, Oscar},
  year      = {2008},
  pages     = {209--218},
  publisher = {IEEE}
}

@inproceedings{mcinnes_2017,
  title     = {Accelerated Hierarchical Density Based Clustering},
  note      = {\doi{10.1109/ICDMW.2017.12}},
  booktitle = {International Conference on Data Mining Workshops ({ICDMW})},
  author    = {McInnes, Leland and Healy, John},
  year      = {2017},
  pages     = {33--42},
  publisher = {IEEE}
}

@article{mcinnes_2018,
  title   = {{UMAP}: Uniform Manifold Approximation and Projection},
  volume  = {3},
  note    = {\doi{10.21105/joss.00861}},
  journal = {Journal of Open Source Software ({JOSS})},
  author  = {McInnes, Leland and Healy, John and Saul, Nathaniel and Großberger, Lukas},
  year    = {2018},
  pages   = {1--2}
}

@inproceedings{misiak_2018,
  title     = {{IslandViz}: A Tool for Visualizing Modular Software Systems in Virtual Reality},
  note      = {\doi{10.1109/VISSOFT.2018.00020}},
  booktitle = {Working Conference on Software Visualization ({VISSOFT})},
  author    = {Misiak, Martin and Schreiber, Andreas and Fuhrmann, Arnulph and Zur, Sascha and Seider, Doreen and Nafeie, Lisa},
  year      = {2018},
  pages     = {112--116},
  publisher = {IEEE}
}

@misc{octoverse_2024,
  title        = {Octoverse},
  url          = {https://octoverse.github.com/},
  journal      = {The GitHub Blog},
  howpublished = {The GitHub Blog},
  author       = {{GitHub}},
  year         = {2024}
}

@preprint{openai_2025,
  title         = {{gpt-oss-120b} \& {gpt-oss-20b} Model Card},
  author        = {OpenAI},
  year          = {2025},
  eprint        = {2508.10925},
  archiveprefix = {arXiv},
  note          = {\doi{10.48550/arXiv.2508.10925}}
}

@inproceedings{raglianti_2023,
  title     = {On the {Rise} of {Modern} {Software} {Documentation}},
  volume    = {263},
  note      = {\doi{10.4230/LIPIcs.ECOOP.2023.43}},
  booktitle = {European Conference on Object-Oriented Programming ({ECOOP})},
  publisher = {Dagstuhl},
  author    = {Raglianti, Marco and Nagy, Csaba and Minelli, Roberto and Lin, Bin and Lanza, Michele},
  year      = {2023},
  pages     = {43:1--43:24}
}

@incollection{runeson_2020,
  title     = {The Design Science Paradigm as a Frame for Empirical Software Engineering},
  isbn      = {978-3-030-32489-6},
  booktitle = {Contemporary Empirical Methods in Software Engineering},
  publisher = {Springer},
  author    = {Runeson, Per and Engström, Emelie and Storey, Margaret-Anne},
  year      = {2020},
  note      = {\doi{10.1007/978-3-030-32489-6_5}},
  pages     = {127--147}
}

@article{schindler_2022,
  title   = {Nested Papercrafts for Anatomical and Biological Edutainment},
  volume  = {41},
  note    = {\doi{10.1111/cgf.14561}},
  journal = {Computer Graphics Forum},
  author  = {Schindler, Marwin and Korpitsch, Thorsten and Raidou, Renata Georgia and Wu, Hsiang-Yun},
  year    = {2022},
  pages   = {541--553}
}

@inproceedings{siddiq_2022,
  address   = {Pittsburgh Pennsylvania},
  title     = {{BERT}-Based {GitHub} Issue Report Classification},
  note      = {\doi{10.1145/3528588.3528660}},
  booktitle = {International Workshop on Natural Language-Based Software Engineering},
  publisher = {ACM},
  author    = {Siddiq, Mohammed Latif and Santos, Joanna C. S.},
  year      = {2022},
  pages     = {33--36}
}

@article{storey_2020,
  title   = {The Who, What, How of Software Engineering Research: {A} Socio-Technical Framework},
  volume  = {25},
  note    = {\doi{10.1007/s10664-020-09858-z}},
  number  = {5},
  journal = {Empirical Software Engineering},
  author  = {Storey, Margaret-Anne and Ernst, Neil A. and Williams, Courtney and Kalliamvakou, Eirini},
  year    = {2020},
  pages   = {4097--4129}
}

@article{tantisuwankul_2019,
  title   = {A Topological Analysis of Communication Channels for Knowledge Sharing in Contemporary {GitHub} Projects},
  volume  = {158},
  note    = {\doi{10.1016/j.jss.2019.110416}},
  journal = {Journal of Systems and Software},
  author  = {Tantisuwankul, Jirateep and Nugroho, Yusuf Sulistyo and Kula, Raula Gaikovina and Hata, Hideaki and Rungsawang, Arnon and Leelaprute, Pattara and Matsumoto, Kenichi},
  year    = {2019},
  pages   = {1--12}
}

@article{venigalla_2024,
  title   = {An Exploratory Study of Software Artifacts on {GitHub} from the Lens of Documentation},
  volume  = {169},
  note    = {\doi{10.1016/j.infsof.2024.107425}},
  journal = {Information and Software Technology},
  author  = {Venigalla, Akhila Sri Manasa and Chimalakonda, Sridhar},
  year    = {2024},
  pages   = {1--21}
}

@inproceedings{warrick_2022,
  title     = {The {OCEAN} Mailing List Data Set: {Network} Analysis Spanning Mailing Lists and Code Repositories},
  note      = {\doi{10.1145/3524842.3528479}},
  booktitle = {International Conference on Mining Software Repositories ({MSR})},
  publisher = {ACM},
  author    = {Warrick, Melanie and Rosenblatt, Samuel F. and Young, Jean-Gabriel and Casari, Amanda and Hébert-Dufresne, Laurent and Bagrow, James},
  year      = {2022},
  pages     = {338--342}
}

@article{weigand_2021,
  title     = {An Artifact Ontology for Design Science Research},
  volume    = {133},
  note      = {\doi{10.1016/j.datak.2021.101878}},
  journal   = {Data \& Knowledge Engineering},
  author    = {Weigand, Hans and Johannesson, Paul and Andersson, Birger},
  year      = {2021},
  pages     = {1--19},
  publisher = {Elsevier}
}

@article{wu_2024,
  title   = {A Survey on Neural Topic Models: Methods, Applications, and Challenges},
  volume  = {57},
  note    = {\doi{10.1007/s10462-023-10661-7}},
  journal = {Artificial Intelligence Review},
  author  = {Wu, Xiaobao and Nguyen, Thong and Luu, Anh Tuan},
  year    = {2024},
  pages   = {1--30}
}

@preprint{zhang_2025,
  title         = {Qwen3 Embedding: Advancing Text Embedding and Reranking Through Foundation Models},
  author        = {Zhang, Yanzhao and Li, Mingxin and Long, Dingkun and Zhang, Xin and Lin, Huan and Yang, Baosong and Xie, Pengjun and Yang, An and Liu, Dayiheng and Lin, Junyang and Huang, Fei and Zhou, Jingren},
  year          = {2025},
  archiveprefix = {arXiv},
  publisher     = {arXiv},
  eprint        = {2506.05176},
  note          = {\doi{10.48550/arXiv.2506.05176}}
}

@inproceedings{zhao_2021,
  title     = {Topic Modelling Meets Deep Neural Networks: {A} Survey},
  volume    = {5},
  note      = {\doi{10.24963/ijcai.2021/638}},
  booktitle = {International Joint Conference on Artificial Intelligence ({IJCAI})},
  author    = {Zhao, He and Phung, Dinh and Huynh, Viet and Jin, Yuan and Du, Lan and Buntine, Wray},
  year      = {2021},
  pages     = {4713--4720},
  publisher = {IJCAI}
}

@inproceedings{mteb,
  title     = {{MTEB}: Massive Text Embedding Benchmark},
  author    = {Muennighoff, Niklas and Tazi, Nouamane and Magne, Loic and Reimers, Nils},
  booktitle = {Conference of the European Chapter of the Association for Computational Linguistics (EACL)},
  year      = {2023},
  publisher = {ACL},
  url       = {https://huggingface.co/spaces/mteb/leaderboard},
  note      = {\doi{10.18653/v1/2023.eacl-main.148}},
  pages     = {2014--2037}
}

\end{document}